# A new efficient staggered grid finite difference scheme for elastic wave equation modeling

Wenquan Liang[1], Chaofan Wu[1], Yanfei Wang[2*], Changchun Yang[2], Xiaobi Xie[3]

**Abstract**

Staggered grid finite difference scheme is widely used for the first order elastic wave equation, which constitutes the basis for least-squares reverse time migration and full waveform inversion. It is of great importance to improve the efficiency and accuracy of wave equation modeling. Usually the same staggered grid finite difference scheme is used for all the spatial derivatives in the first order elastic wave equation. In this paper, we propose a new staggered grid finite difference scheme which can improve the efficiency while preserving the same accuracy for the first order elastic wave equation simulation. It uses second order staggered grid finite difference scheme for some of the first order spatial derivatives while utilizing longer staggered grid finite difference operator for other first order spatial derivatives. We The staggered grid finite difference coefficients of the new finite difference scheme are determined in the space domain by a linear method. We demonstrate by dispersion analysis and numerical simulation the effectiveness of the proposed method.

## 1. Introduction

Staggered grid finite-difference (FD) methods are widely used for first order elastic wave equation modeling because of their high computational efficiency, smaller memory requirement and easy implementation (Virieux 1984,1986; Robertsson et al., 1994; Liu and Sen, 2011a,2011b; Chu and Stoffa, 2012; Wang et al, 2014; Bohlen et al. 2016; Etemadsaeed et al. 2016). They also constitute the basis for full waveform inversion and least squares reverse time migration (LSRTM) (Yang et al, 2016; Ren et al,2017).

Many efforts are paid to improving the simulation efficiency and reducing the grid dispersion of the FD method for the elastic wave equation modeling. Ren and Liu (2014) developed a novel optimal time-space domain staggered grid FD scheme and used least-squares method to get the FD coefficients. Chen et al (2017) used a K-space operator based high order staggered grid FD method to improve simulation accuracy. Yong et al (2017) proposed using optimized equivalent staggered grid FD method with three sets of FD coefficients to improve accuracy.

However, the same staggered grid FD operator length are usually used for all the spatial derivatives in the previous methods. We propose to use different staggered grid FD scheme for the first order spatial derivatives in the first order elastic wave equation. For some of the first order spatial derivatives in the first order elastic wave equation, we only use the simplest second order staggered grid FD scheme. For the other first order spatial derivatives in the elastic wave equation, we use the staggered grid FD scheme with increased operator length. Dispersion analysis and numerical simulation demonstrate that the new staggered grid FD scheme can reduce the numerical simulation time while still preserving high accuracy. Because half of the spatial derivatives are discretized with the simplest second order staggered grid FD scheme, our proposed method can reduce 40 percent (which is related to the length of the operator length) of the simulation time compared with the previous staggered grid FD schemes for the first order elastic wave equation.

[1] College of Resource Engineering, Longyan University, Longyan 364000, People's Republic of China

[2] Key Laboratory of Petroleum Resources Research, Institute of Geology and Geophysics, Chinese Academy of Sciences, Beijing 100029, People's Republic of China. Corresponding author. email:yfwang@mail.iggcas.ac.cn

[3] Institute of Geophysics and Planetary Physics, University of California, Santa Cruz, CA 95064, USA

## 2. Determining the FD coefficients

The first order elastic wave equations in 2D heterogeneous media are (Virieux 1986)

$$\frac{\partial v_x}{\partial t} = \frac{\partial \tau_{xx}}{\partial x} + \frac{\partial \tau_{xz}}{\partial z}, \tag{1}$$

$$\frac{\partial v_z}{\partial t} = \frac{\partial \tau_{xz}}{\partial x} + \frac{\partial \tau_{zz}}{\partial z}, \tag{2}$$

$$\frac{\partial \tau_{xx}}{\partial t} = \alpha^2 \frac{\partial v_x}{\partial x} + (\alpha^2 - 2\beta^2)\frac{\partial v_z}{\partial z}, \tag{3}$$

$$\frac{\partial \tau_{zz}}{\partial t} = \alpha^2 \frac{\partial v_z}{\partial z} + (\alpha^2 - 2\beta^2)\frac{\partial v_x}{\partial x}, \tag{4}$$

$$\frac{\partial \tau_{xz}}{\partial t} = \beta^2 (\frac{\partial v_x}{\partial z} + \frac{\partial v_z}{\partial x}). \tag{5}$$

where $(v_x, v_z)$ is the velocity vector, $(\tau_{xx}, \tau_{zz}, \tau_{xz})$ is the stress vector, $\alpha$ and $\beta$ are the *P*- and *S*- wave propagation speeds respectively. Usually, the identical staggered grid FD scheme are used for all the spatial derivatives in the first order elastic wave equations (1) -(5). Different with the previous methods, we propose to use the simplest second order staggered grid FD scheme for spatial derivatives in equations (3) -(5). Then we substitute equations (3) -(5) into equations (1) and (2), and obtain

$$\frac{\partial^2 v_x}{\partial t^2} = \frac{\alpha^2 \partial(v_{x0,1/2} - v_{x0,-1/2})}{\partial x} + \frac{\beta^2 \partial\left[(v_{x1/2,0} - v_{x-1/2,0})\right]}{\partial z}$$
$$+ \frac{(\alpha^2 - 2\beta^2)\partial(v_{z1/2,0} - v_{z-1/2,0})}{\partial x} + \frac{\beta^2 \partial\left[(v_{z0,1/2} - v_{z0,-1/2})\right]}{\partial z} \qquad (6)$$

$$\frac{\partial^2 v_z}{\partial t^2} = \frac{\alpha^2 \partial(v_{z1/2,0} - v_{z-1/2,0})}{\partial z} + \frac{\beta^2 \partial(v_{z0,1/2} - v_{z0,-1/2})}{\partial x}$$
$$+ \frac{(\alpha^2 - 2\beta^2)\partial(v_{x0,1/2} - v_{x0,-1/2})}{\partial z} + \frac{\beta^2 \partial\left[(v_{x1/2,0} - v_{x-1/2,0})\right]}{\partial x} \qquad (7)$$

where

$$P_{m,j}^n = P(z + mh, x + jh, t + n\tau); P = v_x, v_z \qquad (8)$$

We use increased length staggered grid FD operator for the spatial derivatives in equations (6) and (7). Then we get,

$$\frac{\partial^2 v_x}{\partial t^2} = \frac{\alpha^2}{h^2}\sum_{m=1}^{M} c_m \left[v_{x(0,m)} - v_{x(0,-m+1)} - (v_{x(0,m-1)} - v_{x(0,-m)})\right]$$
$$+ \frac{\beta^2}{h^2}\sum_{m=1}^{M} c_m \left[v_{x(m,0)} - v_{x(-m+1,0)} - (v_{x(m-1,0)} - v_{x(-m,0)})\right]$$
$$+ \frac{\beta^2}{h^2}\sum_{m=1}^{M} c_m \left[v_{z(m-1/2,1/2)} - v_{z(-m+1/2,1/2)} - (v_{z(m-1/2,-1/2)} - v_{z(-m+1/2,-1/2)})\right]$$
$$+ \frac{(\alpha^2 - 2\beta^2)}{h^2}\sum_{m=1}^{M} c_m \left[v_{z(1/2,m-1/2)} - v_{z(1/2,-m+1/2)} - (v_{z(-1/2,m-1/2)} - v_{z(-1/2,-m+1/2)})\right] \qquad (9)$$

and

$$\frac{\partial^2 v_z}{\partial t^2} = \frac{\beta^2}{h^2}\sum_{m=1}^{M} c_m \left[v_{z(0,m)} - v_{z(0,-m+1)} - (v_{z(0,m-1)} - v_{z(0,-m)})\right]$$
$$+ \frac{\alpha^2}{h^2}\sum_{m=1}^{M} c_m \left[v_{z(m,0)} - v_{z(-m+1,0)} - (v_{z(m-1,0)} - v_{z(-m,0)})\right]$$
$$+ \frac{(\alpha^2 - 2\beta^2)}{h^2}\sum_{m=1}^{M} c_m \left[v_{x(m-1/2,1/2)} - v_{x(-m+1/2,1/2)} - (v_{x(m-1/2,-1/2)} - v_{x(-m+1/2,-1/2)})\right]$$
$$+ \frac{\beta^2}{h^2}\sum_{m=1}^{M} c_m \left[v_{x(1/2,m-1/2)} - v_{x(1/2,-m+1/2)} - (v_{x(-1/2,m-1/2)} - v_{x(-1/2,-m+1/2)})\right]$$
$$(10)$$

where *M* is the length of the finite difference operators, $c_m$ are the finite difference coefficients, *h* is the space grid size.

For the second order spatial derivatives in equations (9) and (10), we get the following dispersion relation,

$$2\sum_{m=1}^{M}c_m[\cos(mkh)-\cos((m-1)kh)]=-4\sin(0.5kh)\sum_{m=1}^{M}c_m\sin((m-0.5)kh)=-k^2h^2. \quad (11)$$

For the mixed order spatial derivatives in equations (9) and (10), we get the following dispersion relation,

$$4\sin(0.5k_xh)\sum_{m=1}^{M}c_m\sin((m-0.5)k_zh)=k_xk_z; \quad 4\sin(0.5k_zh)\sum_{m=1}^{M}c_m\sin((m-0.5)k_xh)=k_xk_z. \quad (12)$$

Any of the three equations in (11)-(12) can be used to determine the staggered grid FD coefficient. We use the dispersion relation in equation (11) to determine the staggered grid FD coefficient.

We determine the upper limit of the wavenumber range used for calculating the FD coefficients based on the source frequency, the space grid interval and the S wave velocity (Liang et al., 2013)

$$r=\frac{k_{fd}}{k_{total}}=\frac{2\pi f/v}{\pi/h}=\frac{f_{max}}{(v/2h)}. \quad (13)$$

There are many methods can be used to determine the staggered grid FD coefficient in equation (11) (Etgen, 2007; Liu and Sen, 2011a; Zhang and Yao, 2013). The dispersion relation in equation (11) is linear, therefore we determine the staggered grid FD coefficients by our previously proposed linear method (Liang et al., 2013). It is compared by our previous work that the linear method can determine the FD coefficient almost as good as the optimized method and with the linear method it saves time to determine the FD coefficient (Liang et al., 2015). In the following, we refer the FD scheme with all the spatial derivatives has the same FD operator length as the traditional staggered grid FD scheme. we refer the FD scheme which has different FD operator length for different spatial derivatives as the new staggered grid FD scheme. We refer the staggered grid FD coefficient obtained in the space domain by the Taylor expansion method as the traditional staggered grid FD coefficient (Chu and Stoffa, 2012).

**3 Dispersion analysis**

In this section, we compare the dispersion error of the traditional staggered grid FD scheme and the new staggered grid FD scheme. The dispersion error $\delta$ of the new staggered grid FD scheme for the second order spatial derivative is defined as

$$\delta = 4\sin(0.5kh)\sum_{m=1}^{M} c_m \sin((m-0.5)kh) - k^2 h^2. \tag{14}$$

The dispersion error $\delta$ of the traditional staggered grid FD scheme for the second order spatial derivative is defined as

$$\delta = \left[2\sum_{m=1}^{M} c_m \sin((m-0.5)kh)\right]^2 - k^2 h^2 \tag{15}$$

From figure 1(a) and (b), we get the conclusion that for the second order spatial derivatives, the new staggered grid FD scheme's performance and the traditional staggered grid FD scheme's performance are almost identical.

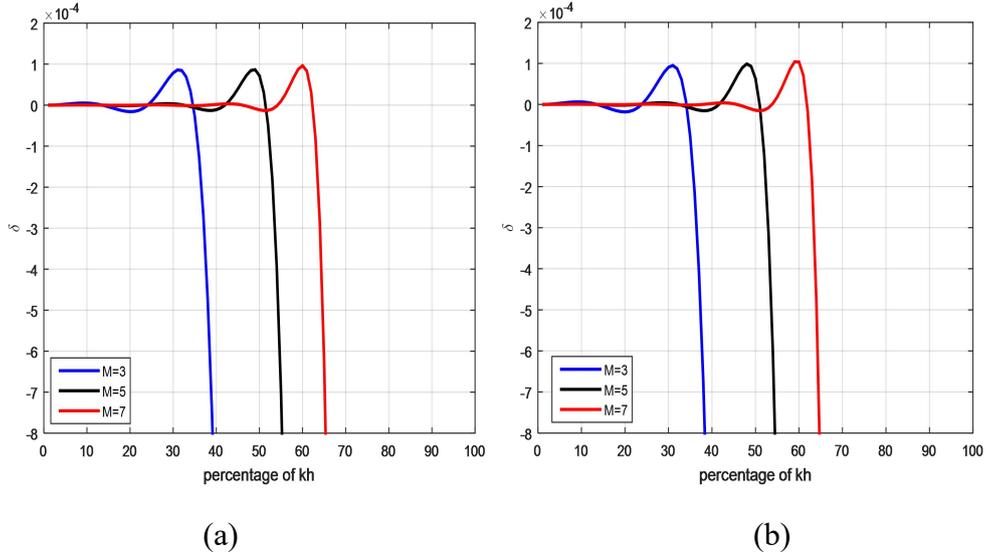

(a)　　　　　　　　　　　　　　(b)

Figure 1 Dispersion error curves for the second order spatial derivatives with different finite difference schemes (a)the traditional staggered grid FD scheme; (b) the new staggered grid FD scheme.

The dispersion error $\varepsilon$ of the new staggered grid FD scheme for the mixed order spatial derivatives is defined as

$$\varepsilon = 2\sin(0.5k_x h)\sum_{m=1}^{M} 2c_m \sin((m-0.5)k_z h) - k_x k_z h^2. \tag{16}$$

The dispersion error $\varepsilon$ of the traditional staggered grid FD scheme for the mixed order spatial derivatives is defined as

$$\varepsilon = \left[\sum_{m=1}^{M} 2c_m \sin((m-0.5)k_x h)\right]\left[\sum_{m=1}^{M} 2c_m \sin((m-0.5)k_z h)\right] - k_x k_z h^2 \quad (17)$$

From figure 2(a) and (b), we get the conclusion that for the mixed order spatial derivatives, the new staggered grid FD scheme is less accurate compared with the traditional staggered grid FD scheme. However, we will find in numerical simulation that the although the new staggered grid FD scheme is less accurate for the mixed order spatial derivatives, its accuracy is still really good.

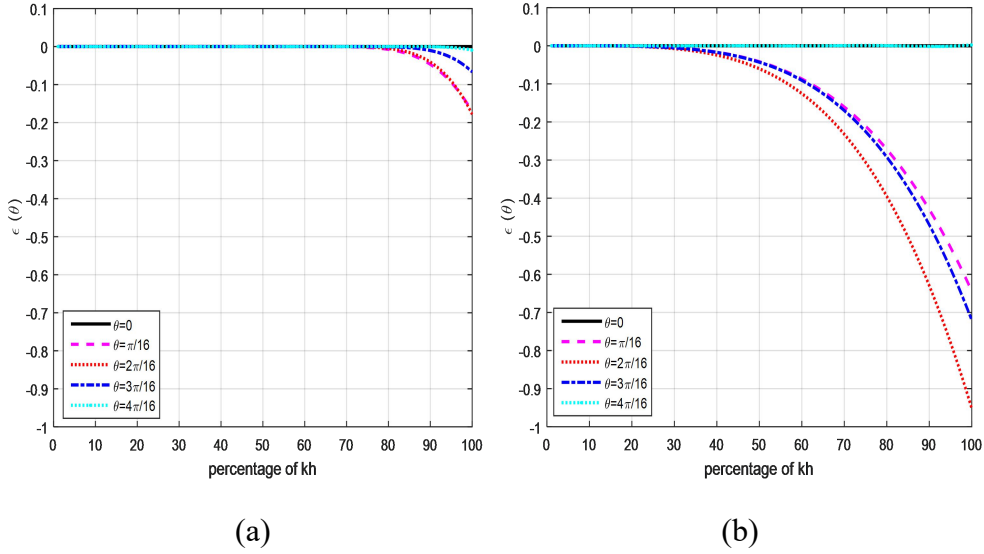

(a)        (b)

Figure 2 Dispersion error curves for the mixed order spatial derivatives with different FD schemes (a) the traditional staggered grid FD scheme; (b) the new staggered grid FD scheme.

## 4 Numerical simulation examples

### 4.1 Numerical modeling in the homogeneous media

We first consider a homogeneous model. The P wave propagation speed is 2598 m/s and the S wave propagation speed is 1500 m/s. The seismic source position is at the center of the model and added to the *Vx* component. The grid space interval is 20m , the time step is 1 ms and the operator length *M* is 7. A Ricker wavelet with the main frequency as 14.3 Hz was used as the seismic source.

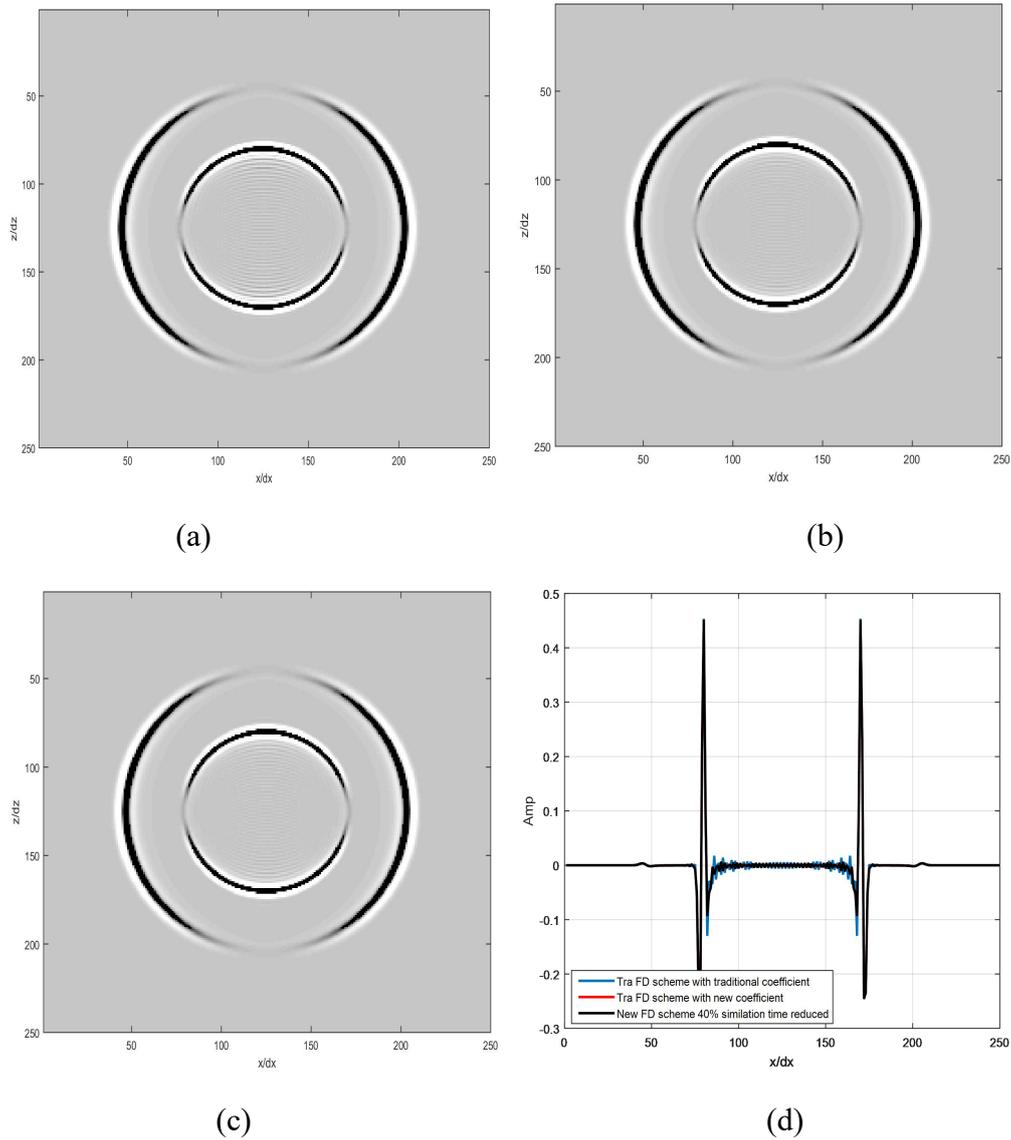

Figure 3 Snapshots and slices of snapshots of the horizontal component at 698 ms obtained by different simulation methods (a)the traditional staggered gird FD scheme with traditional FD coefficient; (b)the traditional staggered grid FD scheme with new FD coefficient; (c) the new staggered grid FD scheme with new FD coefficient; (d)slices of snapshots at x/dx=125.

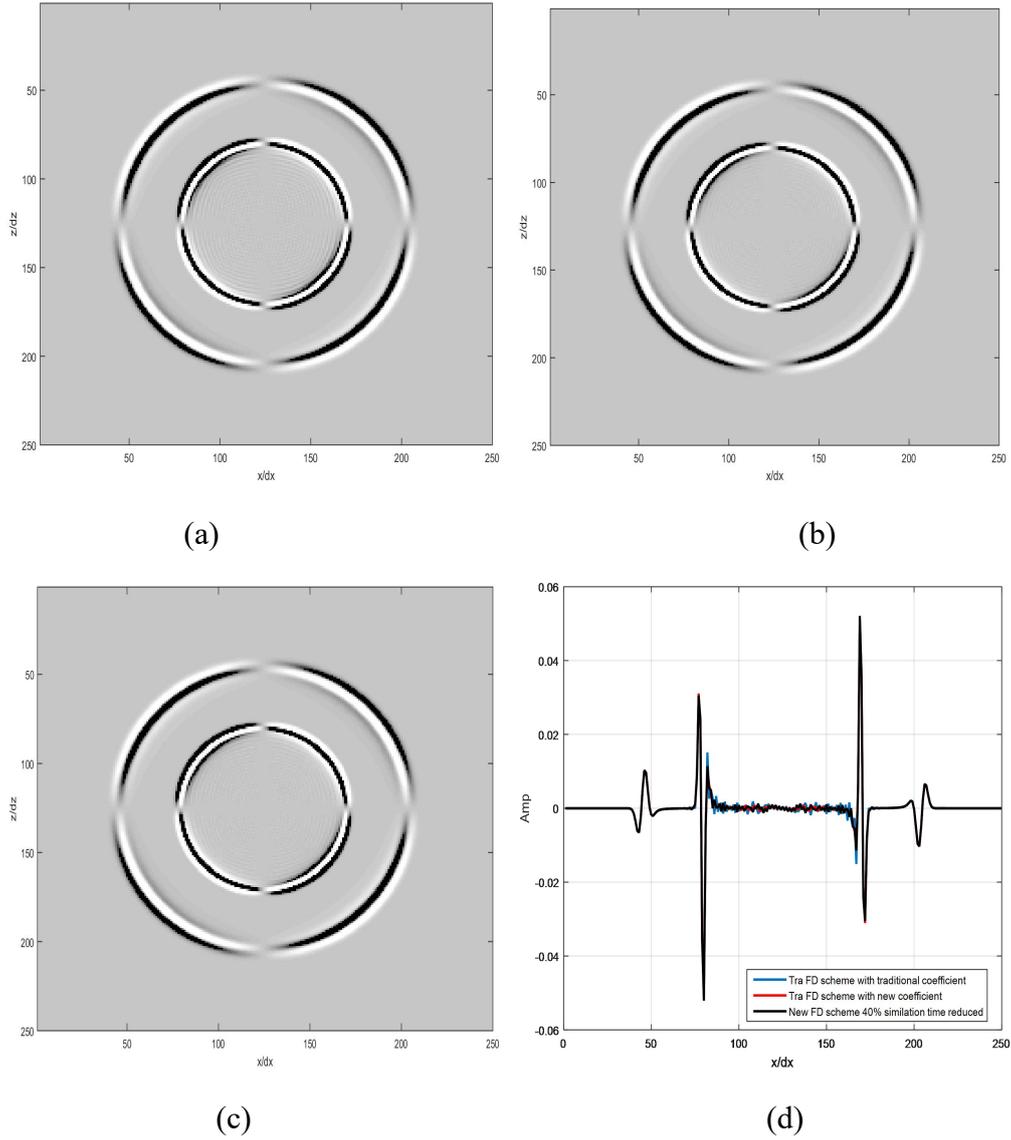

(a)　　　　　　　　　　　　　　(b)

(c)　　　　　　　　　　　　　　(d)

Figure 4 Snapshots and slices of snapshots of the vertical component at 698 ms obtained by different simulation methods. (a)the traditional staggered gird FD scheme with traditional FD coefficient; (b)the traditional staggered grid FD scheme with new FD coefficient; (c) the new staggered grid FD scheme with new FD coefficient; (d)slices of snapshots at z/dz=120.

The snapshots of the horizontal component obtained by different staggered grid FD methods are presented in figure 3(a)-(c). Figure 3(a) is obtained with the traditional staggered grid FD scheme with the traditional FD coefficient. The grid dispersion is obvious. Figure 3(b) is obtained with the traditional staggered grid FD scheme with the new FD coefficient. Compared with figure 3(a), the grid dispersion is suppressed. Figure 3(c) is obtained with the new staggered grid FD scheme. The grid dispersion in figure 3(b) and (c) are very similar, which is further demonstrated in figure 3(d). The same pattern can be observed from figure 4(a)-(d). However, with the new staggered grid FD scheme, we can save about 40 percent of the simulation time.

## 4.2 Salt model

Figure 5 shows the salt model from Society of Exploration of Geophysics. The S wave velocity is obtained from the P wave velocity. The position of the seismic source is plotted as a red star. The spatial sampling interval is 12.5 m, temporal step is 1 ms and $M = 7$ for the staggered grid FD operators.

Figure 6 are the seismic records of the horizontal component obtained by different FD methods. Figure 6(a) is obtained with the traditional FD scheme with the traditional staggered grid FD coefficient. The grid dispersion is severe. Figure 6(b) is obtained the with the traditional FD scheme with the staggered grid FD coefficient obtained by the linear method. Figure 6(c) is obtained the with the new FD scheme with the staggered grid FD coefficient obtained by the linear method. It is observed that the grid dispersion in figure 6(b) and 6(c) are smaller than the grid dispersion in figure 6(a). Figure 6(d) are seismograms obtained from figure 6(a)-(c). It further demonstrated that the grid dispersion in figure 6(b) and (c) are similar to each other and smaller than the grid dispersion in figure 6(a). However, with the new FD scheme, the simulation time is reduced by more than 40 percent. In our simulation, there are 565 grids in the z direction and 890 grids in the x direction. With the traditional staggered grid FD scheme, the simulation time is 920 seconds. With the new staggered grid FD grid scheme, the simulation time is 530 seconds. The huge reduction in simulation time is mainly because we use shorter staggered grid FD operator for the first order spatial derivatives in equations (3), (4) and (5). Figure 7 are the seismic records of the vertical component obtained by different FD methods. The same pattern can be observed from figure 7(a)-(d).

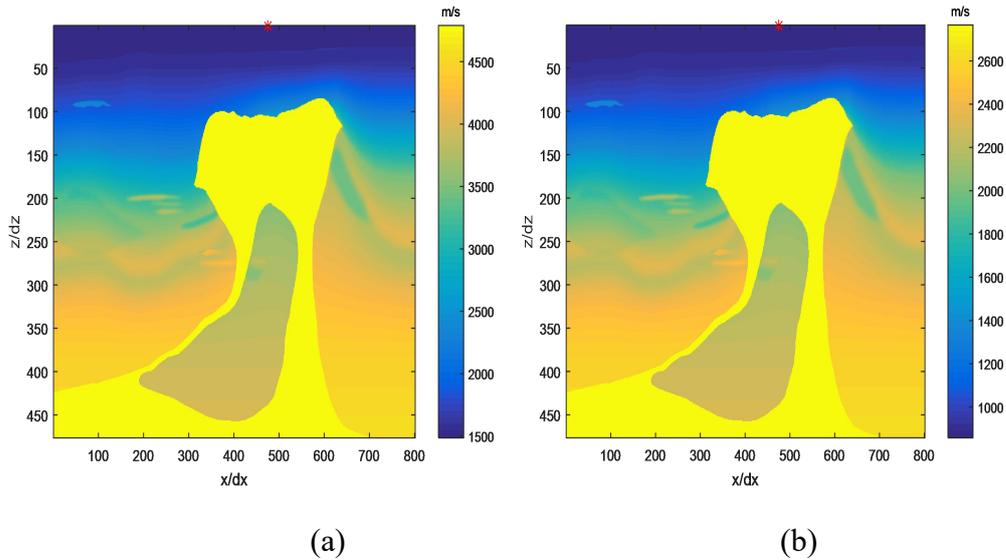

(a)                    (b)

Figure 5 BP salt model. (a) P wave velocity; (b) S wave velocity

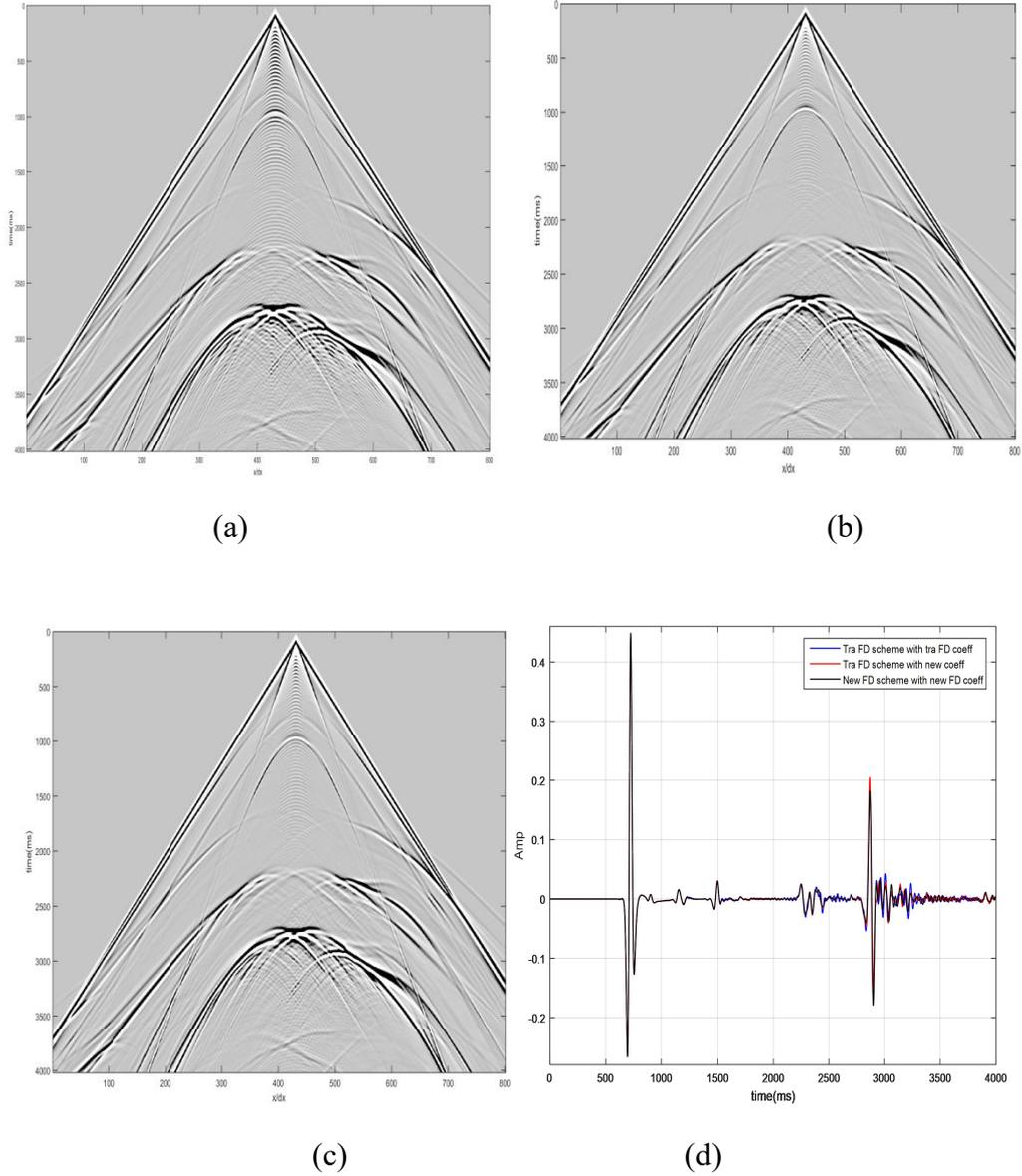

Figure 6 Seismic records of the horizontal component obtained with different methods. (a)the traditional FD scheme with traditional staggered grid FD coefficient; (b) the traditional FD scheme with new staggered grid FD coefficient; (c) the new FD scheme with new staggered grid FD coefficient; (d)seismograms obtained from (a)-(c) at position x/dx=355.

## 5 Conclusion

We proposed a new staggered grid FD scheme to reduce the simulation time while preserve almost the same simulation accuracy for the first order elastic wave equation. Through dispersion analysis and numerical simulation, we conclude that our new staggered grid FD scheme is more efficient while preserving the high accuracy. As a result, our method can be a substitute for the other staggered grid FD schemes used in first order elastic wave equation extrapolation, which are essential in forward seismic wave modeling and reverse-time migration.

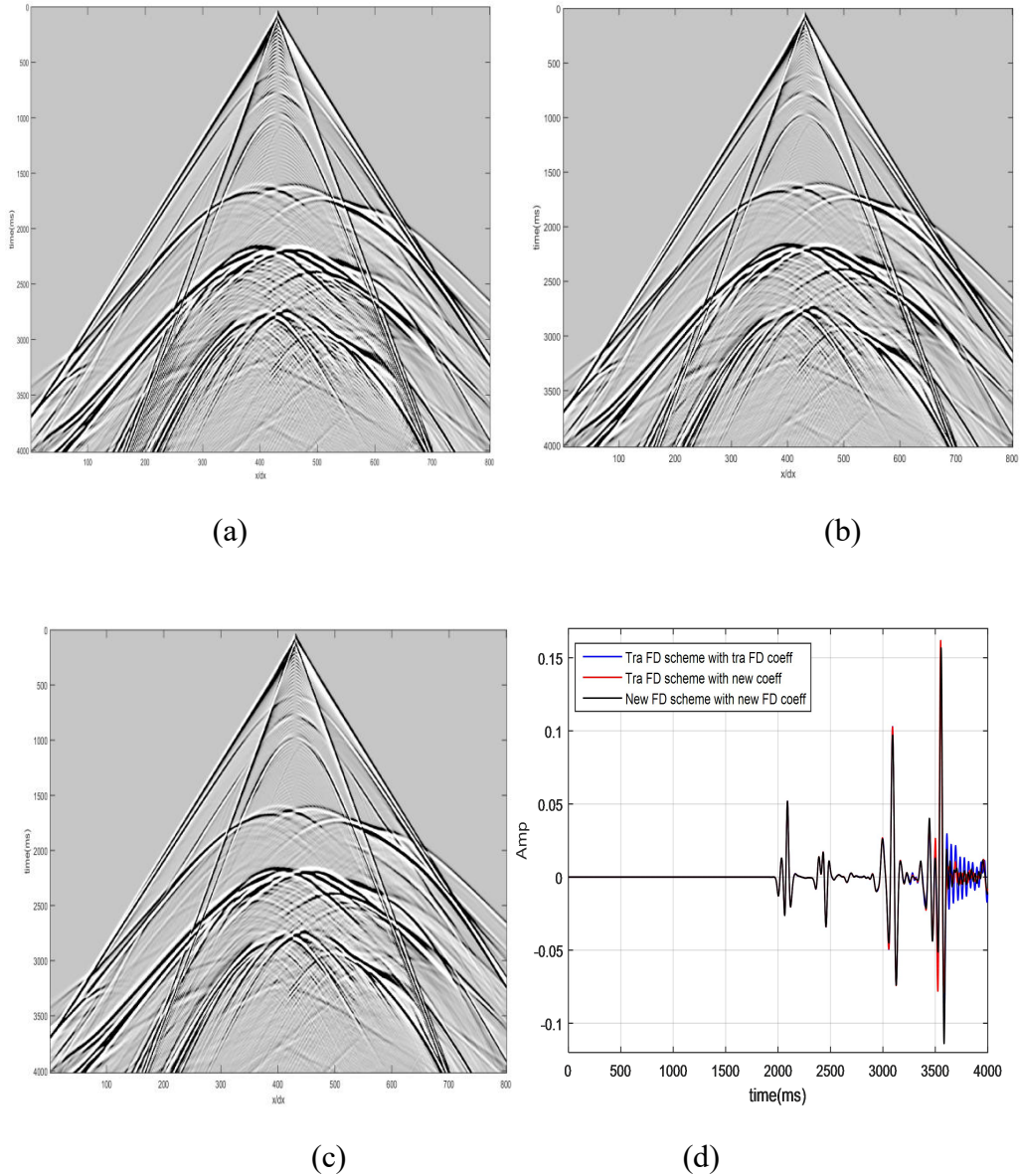

Figure 7 Seismic records of the vertical component obtained with different methods. (a)the traditional FD scheme with traditional staggered grid FD coefficient; (b) the traditional FD scheme with new staggered grid FD coefficient; (c) the new FD scheme with new staggered grid FD coefficient; (d) seismograms obtained from (a)-(c) at position x/dx=200.

**References**

Bohlen, T., Wittkamp, F. 2016.Three-dimensional viscoelastic time-domain finite-difference seismic modelling using the staggered Adams-Bashforth time integrator. Geophysical Journal International, 204(3), 1781-1788.

Chen, H., Zhou, H., Zhang, Q., Chen, Y. 2017. Modeling elastic wave propagation using *K* space operator-based temporal high-order staggered-grid finite-difference method. IEEE Transactions on Geoscience and Remote Sensing, 55(2), 801-815.

Chu, C., Stoffa, P. L. 2012. Determination of finite-difference weights using scaled binomial windows. Geophysics, *77*(3), W17-W26.

Etemadsaeed, L., Moczo, P., Kristek, J., Ansari, A., Kristekova, M. 2016. A no-cost improved velocity–stress staggered-grid finite-difference scheme for modelling seismic wave propagation. Geophysical Journal International, 207(1), 481-511.

Etgen, J. T., 2007, A tutorial on optimizing time domain finite-difference scheme: "Beyond Holberg": Stanford Exploration Project Report, 129, 33-43.

Liang, W. Q., Yang, C. C., Wang, Y. F., Liu H W 2013. Acoustic wave equation modeling with new time-space domain finite difference operators. Chinese Journal of Geophysics, 56(6), 840-850.

Liu, Y., Sen, M. K. 2011a. Finite-difference modeling with adaptive variable-length spatial operators. Geophysics, 76(4), T79-T89.

Liu, Y., Sen, M. K. 2011b. Scalar wave equation modeling with time-space domain dispersion-relation-based staggered-grid finite-difference schemes. Bulletin of the Seismological Society of America, 101(1), 141-159.

Ren, Z., Liu, Y. 2014. Acoustic and elastic modeling by optimal time-space-domain staggered-grid finite-difference schemes. Geophysics, 80(1), T17-T40.

Ren, Z., Liu, Y., Sen, M. K. 2017. Least-squares reverse time migration in elastic media. Geophysical Journal International, 208(2), 1103-1125.

Robertsson, J. O., Blanch, J. O., Symes, W. W. 1994. Viscoelastic finite-difference modeling. Geophysics, 59(9), 1444-1456.

Virieux, J. 1984. SH-wave propagation in heterogeneous media: Velocity-stress finite-difference method. Geophysics, 49(11), 1933-1942.

Virieux, J. 1986. P-SV wave propagation in heterogeneous media: Velocity-stress finite-difference method. Geophysics, 51(4), 889-901.

Wang, Y., Liang, W., Nashed, Z., Li, X., Liang, G., Yang, C. 2014. Seismic modeling by optimizing regularized staggered-grid finite-difference operators using a time-space-domain dispersion- relationship-preserving method. Geophysics, 79(5), T277-T285.

Yang, J., Liu, Y., Dong, L. 2016. Least-squares reverse time migration in the presence of density variations. Geophysics, 81(6), S497-S509.

Yong, P., Huang, J., Li, Z., Liao, W., Qu, L., Li, Q., & Liu, P. (2017). Optimized equivalent staggered-grid FD method for elastic wave modelling based on plane wave solutions. Geophysical Journal International, 208(2), 1157-1172.

Zhang, J.H., Yao, Z.X., 2013, Optimized finite-difference operator for broadband seismic wave modeling: Geophysics, 78, A13-A18.